\documentclass[a4paper, 12pt]{article}
\usepackage[T1, T2A]{fontenc}
\usepackage[english,ukrainian]{babel}

\usepackage{csquotes}
\usepackage[
unicode,
colorlinks,%
linkcolor=blue,citecolor=red,urlcolor=blue,
hyperindex,%
plainpages=false,%
bookmarksopen,%
bookmarksnumbered%
]{hyperref}
\usepackage{geometry}
\usepackage[
    backend=biber,
    style=numeric-comp,
    sorting=none,
    maxbibnames=4,
    minbibnames=4,
    maxcitenames=2,
    mincitenames=1,
    doi=true,
    url=true
]{biblatex}
\renewbibmacro*{editor+others}{%
  \ifnameundef{editor}
    {}
    {
     \printnames{editor}
     \setunit{\addspace}
     \printtext{editors}
     \clearname{editor}
    }
}

\renewbibmacro{in:}{}

\DeclareFieldFormat{pages}{#1}

\DefineBibliographyStrings{ukrainian}{
  andothers = {et\addabbrvspace al\adddot},
}

\DeclareFieldFormat{volume}{\textbf{#1}}
\renewbibmacro*{volume+number+eid}{
    \printfield{volume}
    \iffieldundef{number}{}{\addcomma\printfield{number}}
    \setunit{\addcomma\space}
    \printfield{eid}}

\DeclareFieldFormat{doi}{%
  \newline\url{https://doi.org/#1}}

\usepackage{amsmath}
\usepackage{graphicx}

\usepackage{setspace}
\begin{document}
\sloppy

\begin{flushleft}

{\Large\textbf{Regularization and Chaotization of Maximal Attractors
in the Sommerfeld--Kononenko Non-Ideal
``Spherical Pendulum--Electric Motor''\ System
with Time Delays}}

\textbf{O.O. Horchakov$^{1,2}$,
A.Yu. Shvets$^{1*}$}\\

$^1$ Institute of Mathematics
of the National Academy of Sciences of Ukraine,
3 Tereshchenkivska St.,
Kyiv 01024, Ukraine\\

$^2$ Kyiv School of Economics,
3 Mykola Shpak St.,
Kyiv 03113, Ukraine\\

$^*$Corresponding author\\

e-mail:
\href{mailto:o.horchakov@imath.kiev.ua}
{o.horchakov@imath.kiev.ua}\\

ORCID:
\href{https://orcid.org/0009-0006-3664-8812}
{https://orcid.org/0009-0006-3664-8812}\\

e-mail:
\href{mailto:oshvets@imath.kiev.ua}
{oshvets@imath.kiev.ua}\\

ORCID:
\href{https://orcid.org/0000-0003-0330-5136}
{https://orcid.org/0000-0003-0330-5136}

\end{flushleft}

\vspace{0.4cm}

\begin{flushleft}

The influence of the time-delay parameters
on the bifurcations of ``non-classical''
maximal attractors in the
Sommerfeld--Kononenko non-ideal dynamical system
``spherical pendulum--electric motor''
is investigated.
Regular and chaotic maximal attractors
of this system,
as well as their bifurcations,
are described.
It is established that the presence
of time delays can fundamentally change
the type of limit sets
of the considered dynamical system
and significantly alter the scenarios
of transitions to deterministic chaos.

\noindent\textbf{2020 MSC:}
37G25, 37G35, 37L30, 37M20

\end{flushleft}

\section*{Introduction}

The study of dynamical systems with limited excitation
is an important research area in modern nonlinear dynamics.
Such systems were first considered in the pioneering works
of Sommerfeld \cite{sommerfeld1902, sommerfeld1904}
and Rocard \cite{rocard1943}.
Subsequently, the theoretical foundations of this field
were developed in Kononenko's monograph
\cite{kononenko1969},
where the corresponding axiomatic framework
was introduced and mathematical models
for numerous applied problems were constructed.
In the study of such
``oscillatory subsystem--excitation source'' systems,
it is assumed that the power of the energy source
is comparable with the power consumed
by the oscillatory subsystem.
Under this assumption,
the operating regime of the energy source
depends on the oscillations of the subsystem,
and therefore its action cannot be represented
as a prescribed function of time.
Such energy sources are referred to as
non-ideal.
Dynamical systems of the
``oscillatory subsystem--excitation source'' type
are now commonly called
Sommerfeld--Kononenko non-ideal dynamical systems.

In contrast,
energy sources for which no power limitations
are taken into account
are referred to as ideal.
A large number of dynamical systems have been investigated
for which assuming the excitation source to be ideal
leads to significant errors
in the analysis of the system dynamics.
In particular,
neglecting the non-ideal nature
of the excitation source
may result in an incorrect identification
of the dynamical regime.
For example,
theoretical analysis may predict
a Lyapunov-stable limit cycle
or an equilibrium state,
whereas experimental investigations
reveal deterministic chaos,
and vice versa. The studies reported in \cite{KrasnopolskayaShvets1991,krasnopolskaya1992,krasnopolskaya1993}
demonstrated that deterministic chaos
may arise solely as a consequence
of the nonlinear interaction
between the oscillatory subsystem
and its excitation source.
Among the studies devoted to non-ideal dynamical systems
published during the last fifteen years,
the following works may be highlighted:
\cite{ShvetsMakaseyev2012,Balthazar2018Over,Cveticanin2018,
Mikhlin2020Resonance,Warminski2022Nonlinear,
pintoGenerating2023,Petrocino2023,
donetskyi2023,Lebedenko2025Stationary,SeitDzhelilShvets2026Delay,SeitDzhelilShvets2026}.

\section*{Problem Formulation}

\begin{figure}[htbp]
\begin{minipage}[t]{0.7\linewidth}
\center{\includegraphics[width=0.7\linewidth]{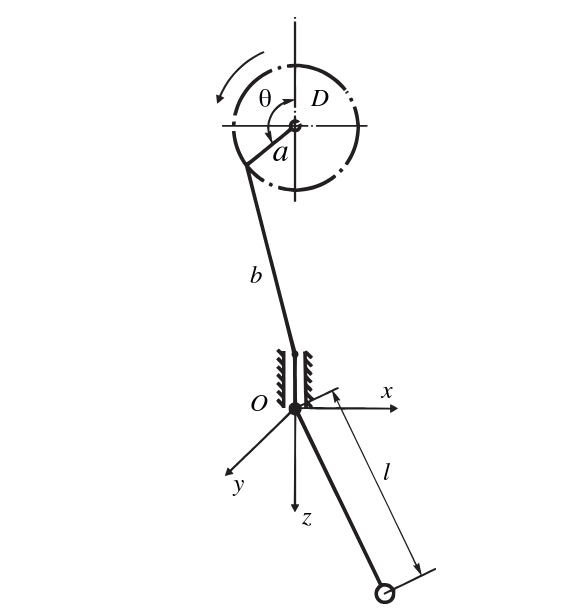}}
\end{minipage}
\caption{``Pendulum--electric motor'' system}
\label{ris:image1}
\end{figure}

Consider a spherical pendulum excited
by an electric motor,
whose schematic diagram
is shown in Fig.~\ref{ris:image1}.
The suspension point of the pendulum
is connected to the motor rotor
through a crank--slider mechanism.

The equations of motion
and the dynamical behavior
of a spherical pendulum
under different types of excitation
of its suspension point
by an ideal energy source
were studied in
\cite{miles1962,miles1984,miles1984Fara}.
The equations of motion
of a spherical pendulum
whose suspension point
is excited vertically
by a limited-power electric motor
were derived in
\cite{krasnopolskaya1992,shvets2007}
in the form
of a fifth-order nonlinear system
of differential equations (\ref{a}).
In this system,
the phase variables
$y_1$, $y_2$, $y_4$, and $y_5$
describe the pendulum motion,
whereas the phase variable
$y_3$
is proportional
to the angular velocity
of the motor shaft.
The variable $\tau$
denotes the dimensionless
``slow'' time.

The parameter $E$
corresponds to the slope
of the static characteristic
of the electric motor,
whereas the parameters
$C$, $D$, and $F$
are multiparameters
that characterize
a number of physical
and geometrical properties
of the
``spherical pendulum--electric motor''
system.
A detailed derivation
of system (\ref{a})
and a complete description
of the parameters
$C$, $D$, and $F$
can be found in
\cite{krasnopolskaya1992}.

\begin{equation}
\begin{aligned}
\frac{dy_1}{d\tau} &= C y_1
- \left[y_3 + \frac{1}{8}\left(y_1^2 + y_2^2 + y_4^2 + y_5^2\right)\right] y_2
 - \frac{3}{4}\left(y_1 y_5 - y_2 y_4\right) 
+ 2y_2, \\[6pt]
\frac{dy_2}{d\tau} &= C y_2
 + \left[y_3 + \frac{1}{8}\left(y_1^2 + y_2^2 + y_4^2 + y_5^2\right)\right] y_1
 - \frac{3}{4}\left(y_1 y_5 - y_2 y_4\right) y_5
 + 2y_1, \\[6pt]
\frac{dy_3}{d\tau} &= D\left(y_1 y_2 + y_4 y_5\right) + E y_3 + F, \\[6pt]
\frac{dy_4}{d\tau} &= C y_4
 - \left[y_3 + \frac{1}{8}\left(y_1^2 + y_2^2 + y_4^2 + y_5^2\right)\right] y_5
 + \frac{3}{4}\left(y_1 y_5 - y_2 y_4\right) y_1
 + 2y_5, \\[6pt]
\frac{dy_5}{d\tau} &= C y_5
 + \left[y_3 + \frac{1}{8}\left(y_1^2 + y_2^2 + y_4^2 + y_5^2\right)\right] y_4
 + \frac{3}{4}\left(y_1 y_5 - y_2 y_4\right) y_2
 + 2y_4.
 \label{a}
\end{aligned}
\end{equation}
Thus, the
``spherical pendulum--electric motor''
system was fundamentally considered
as a Sommerfeld--Kononenko
non-ideal dynamical system.
It should be particularly emphasized
that taking into account
the nonlinear interaction
between the pendulum
and the excitation source
made it possible
to reveal deterministic chaos
in this dynamical system.
It should also be noted
that if the oscillations
of the spherical pendulum
are modeled
under the assumption
of an ideal excitation source,
that is,
by neglecting
the interaction
between the pendulum
and the excitation source,
then,
as established in
\cite{miles1984Fara,Krasnopolskaya1994Resonance},
no chaotic steady-state oscillations
can be detected
when the suspension point
of the pendulum
moves in the vertical direction.

More detailed studies
of the dynamical behavior
of the
``spherical pendulum--electric motor''
system,
performed in
\cite{Shvets2021NewTypes,ShvetsDonetskyi2022_1,donetskyi2023},
revealed the existence
of the so-called
maximal attractors
(Milnor attractors)
\cite{Milnor1985Correction,Anishchenko2014Deterministic}
in this system.
These represent
a highly specific class
of attracting invariant sets
that,
in general,
do not satisfy
the ``classical''
definition of an attractor.
It was established
in
\cite{Shvets2021NewTypes,donetskyi2023}
that,
in the parameter space
of system (\ref{a}),
the equilibrium state
is the only
classical attractor.
All other attractors
of this system
are maximal attractors.
Moreover,
these attracting sets
(maximal attractors)
may be either
regular
or chaotic.

Let us briefly discuss
the differences
between maximal attractors
and classical attractors.
All classical attractors,
including equilibrium states,
limit cycles,
invariant tori,
and various types
of chaotic attractors,
are isolated invariant sets.
That is,
if $\mathcal{V}$
and $\mathcal{W}$
are two distinct attractors,
then
\begin{equation}
    \inf_{\substack{x \in \mathcal{V}\\ z \in \mathcal{W}}}
    \sigma(x,z) > 0,
    \label{new}
\end{equation}
where $\sigma$
denotes a metric,
typically the Euclidean metric,
in the phase space
of the dynamical system.

However, in addition to such
``classical'' attractors,
some dynamical systems
possess invariant sets
that do not satisfy
the isolation condition (\ref{new}).
The invariant sets
of such a dynamical system
can be grouped into families
sharing common properties,
including non-isolation
and an identical spectrum
of Lyapunov characteristic exponents (LCEs).
Among other properties,
such families
may also possess
the attracting property,
which distinguishes
an attractor
from other invariant sets
in the case
of isolated attractors.
For a family
of invariant sets,
the attracting property
manifests itself
in the sense
that every trajectory
originating
from a certain open set
approaches
not the entire family
as a whole,
but one of its members.
Such families
form the so-called
``maximal attractors,''
whose definition
is given in
\cite{Milnor1985Correction,
Anishchenko2014Deterministic,
Shvets2023_Hydrodynamic}.
We emphasize once again
that maximal attractors,
generally speaking,
are not attractors
from the
``classical'' point of view. 

The main objective
of the present study
is to investigate
the influence
of time-delay parameters
on the regular
and chaotic dynamics
of the system
of maximal attractors
in the
``spherical pendulum--electric motor''
system.
Both the delay
in the influence
of the motor shaft motion
on the pendulum oscillations
and the reciprocal delay
in the influence
of the pendulum oscillations
on the shaft rotation
will be considered.
The study
of the influence
of time delays
on the dynamical behavior
of pendulum systems
dates back
to the 1980s. Among the earliest studies devoted to the influence of delay on the stability of the equilibrium states of a physical pendulum whose suspension point moves according to a prescribed law was the paper \cite{Mitropolsky1980,}.
Without presenting
a detailed review
of this research area,
we note
that the presence
of time delays
has been shown,
in some cases,
to stabilize
unstable equilibrium states,
whereas in others
it destabilizes
stable equilibria.
Moreover,
time delays
may lead
to the emergence
of new types
of chaotic attractors
and alter
the scenarios
of transitions
to deterministic chaos.
Taking into account
the physical meaning
of the phase variables
$y_1$, $y_2$, $y_3$, $y_4$, and $y_5$,
the dynamical behavior
of the considered
non-ideal system
with time delays
can be described
by the following system
of differential equations
\begin{equation}
\begin{aligned}
\frac{dy_1(\tau)}{d\tau} &= C y_1(\tau)
 - \left[y_3(\tau-\delta)
 + \frac{1}{8}\bigl(y_1^2(\tau)+y_2^2(\tau)+y_4^2(\tau)+y_5^2(\tau)\bigr)\right] y_2(\tau) \\
&\quad
 - \frac{3}{4}\bigl(y_1(\tau)y_5(\tau)-y_2(\tau)y_4(\tau)\bigr)y_4(\tau)
 + 2y_2(\tau),\\[6pt]
\frac{dy_2(\tau)}{d\tau} &= C y_2(\tau)
 + \left[y_3(\tau-\delta)
 + \frac{1}{8}\bigl(y_1^2(\tau)+y_2^2(\tau)+y_4^2(\tau)+y_5^2(\tau)\bigr)\right] y_1(\tau) \\
&\quad
 - \frac{3}{4}\bigl(y_1(\tau)y_5(\tau)-y_2(\tau)y_4(\tau)\bigr)y_5(\tau)
 + 2y_1(\tau), \\[6pt]
\frac{dy_3(\tau)}{d\tau} &= D\bigl[y_1(\tau-\rho)y_2(\tau-\rho)
 + y_4(\tau-\rho)y_5(\tau-\rho)\bigr]
 + E y_3(\tau) + F, \\[6pt]
\frac{dy_4(\tau)}{d\tau} &= C y_4(\tau)
 - \left[y_3(\tau-\delta)
 + \frac{1}{8}\bigl(y_1^2(\tau)+y_2^2(\tau)+y_4^2(\tau)+y_5^2(\tau)\bigr)\right] y_5(\tau) \\
&\quad
 + \frac{3}{4}\bigl(y_1(\tau)y_5(\tau)-y_2(\tau)y_4(\tau)\bigr)y_1(\tau)
 + 2y_5(\tau), \\[6pt]
\frac{dy_5(\tau)}{d\tau} &= C y_5(\tau)
 + \left[y_3(\tau-\delta)
 + \frac{1}{8}\bigl(y_1^2(\tau)+y_2^2(\tau)+y_4^2(\tau)+y_5^2(\tau)\bigr)\right] y_4(\tau) \\
&\quad
 + \frac{3}{4}\bigl(y_1(\tau)y_5(\tau)-y_2(\tau)y_4(\tau)\bigr)y_2(\tau)
 + 2y_4(\tau). 
 \label{b}
\end{aligned}
\end{equation}
In system (\ref{b}),
$\delta\geq 0$
is the time delay
in the influence
of the motor shaft rotation
on the pendulum oscillations,
whereas
$\rho\geq 0$
is the time delay
in the influence
of the pendulum oscillations
on the rotation
of the motor shaft.

Unlike the five-dimensional
system of equations (\ref{a})
without time delays,
system (\ref{b})
with time delays
is infinite-dimensional.
Even the investigation
of the nonlinear system (\ref{a})
is highly challenging
and can be carried out
in detail
only by means
of numerical methods
and algorithms.
The general methodology
for performing
such numerical computations
is described in
\cite{kuznetsov2006,Skiadas2016,shvets2021pendulum}.
These difficulties
become substantially greater
for the infinite-dimensional
system (\ref{b}).
Therefore,
the first step
toward overcoming them
is to reduce
system (\ref{b})
to a finite-dimensional one.
This can be achieved
rather straightforwardly
by assuming
that the time delays
$\delta$ and $\rho$
are small.
Indeed,
for small values
of the time delays,
the corresponding
delayed phase variables
can be expanded
into Maclaurin series
with respect
to the delay parameters.
Thus,
we may write
\begin{equation}
\begin{aligned}
y_i(\tau-\rho) &= y_i(\tau)-\rho\frac{dy_i(\tau)}{d\tau}+\rho^2\ldots,
\qquad i\in\{1,2,4,5\},\\
y_3(\tau-\delta) &= y_3(\tau)-\delta\frac{dy_3(\tau)}{d\tau}+\delta^2\ldots.
\label{c}
\end{aligned}
\end{equation}
Neglecting
the terms
of order higher
than the first
with respect
to the time delays,
we substitute
expressions (\ref{c})
into system (\ref{b}).
After straightforward
transformations,
we obtain
the following system:
\begin{equation}
\begin{aligned}
\frac{dy_1}{d\tau} &= C y_1
-\Bigl(y_3
-\delta D(y_1y_2+y_4y_5)
-\delta E y_3
-\delta F
+\tfrac18(y_1^2+y_2^2+y_4^2+y_5^2)\Bigr)y_2 \\
&\quad
-\tfrac34(y_1y_5-y_2y_4)y_4
+2y_2, \\[6pt]
\frac{dy_2}{d\tau} &= C y_2
+\Bigl(y_3
-\delta D(y_1y_2+y_4y_5)
-\delta E y_3
-\delta F
+\tfrac18(y_1^2+y_2^2+y_4^2+y_5^2)\Bigr)y_1 \\
&\quad
-\tfrac34(y_1y_5-y_2y_4)y_5
+2y_1, \\[6pt]
\frac{dy_3}{d\tau} &= D\Bigl[
y_1y_2+y_4y_5
-\rho\Bigl(
2C(y_1y_2+y_4y_5) 
+(y_3+\tfrac18(y_1^2+y_2^2+y_4^2+y_5^2)+2)\\ 
&\qquad
\times(y_1^2+y_4^2) -(y_3+\tfrac18(y_1^2+y_2^2+y_4^2+y_5^2)-2)(y_2^2+y_5^2)
\Bigr)\Bigr]
+E y_3+F, \\[6pt]
\frac{dy_4}{d\tau} &= C y_4
-\Bigl(y_3
-\delta D(y_1y_2+y_4y_5)
-\delta E y_3
-\delta F
+\tfrac18(y_1^2+y_2^2+y_4^2+y_5^2)\Bigr)y_5 \\
&\quad
+\tfrac34(y_1y_5-y_2y_4)y_1
+2y_5, \\[6pt]
\frac{dy_5}{d\tau} &= C y_5
+\Bigl(y_3
-\delta D(y_1y_2+y_4y_5)
-\delta E y_3
-\delta F
+\tfrac18(y_1^2+y_2^2+y_4^2+y_5^2)\Bigr)y_4 \\
&\quad
+\tfrac34(y_1y_5-y_2y_4)y_2
+2y_4.
\label{d}
\end{aligned}
\end{equation}

 In system (\ref{d}),
the quantities
$\delta$ and $\rho$
no longer represent
argument time delays,
but are simply
parameters
of the system.
However,
taking into account
the origin
of these parameters,
we shall continue
to refer to them
as time delays.
It should be noted
that the reduction
of the delay system (\ref{b})
to the system
without time delays (\ref{d})
is applicable
only for small values
of the time delays.
If no additional restrictions
are imposed
on the delays,
more sophisticated
reduction techniques
must be employed
\cite{Magnitskii2006, ShvetsMakaseyev2012}.

\section*{The Influence of Time Delays
on Maximal Attractors}

Consider system (\ref{d})
for the following parameter values
\begin{equation}
E=-1.17,\;
C=-0.5,\;
D=-1,\;
F=0.5.
\label{par}
\end{equation}

The time delay $\delta$
will be used
as the bifurcation parameter.
With regard to the time delays,
it is quite natural to assume
that the delay
in the influence
of the motor
on the pendulum oscillations
is equal
to the delay
in the influence
of the pendulum
on the operation
of the motor.
Therefore,
throughout this paper
we assume that
$\rho=2\delta$.

Let us briefly describe
the main numerical methods
and algorithms
used to investigate
the dynamical behavior
of system (\ref{d}).
The solutions
of this system
were computed
using the fourth-order
Runge--Kutta method
\cite{hairer1987}.
The Hénon method
was employed
to construct
bifurcation diagrams
and Poincaré sections
\cite{henon1976}.
The spectrum
of Lyapunov exponents
was computed
using the generalized algorithm
of Benettin \textit{et al.}
\cite{benettin1976,benettin1980}.
 
\begin{figure}[htbp]
\begin{minipage}[htbp]{0.5\linewidth}
\center{\includegraphics[width=1\linewidth]{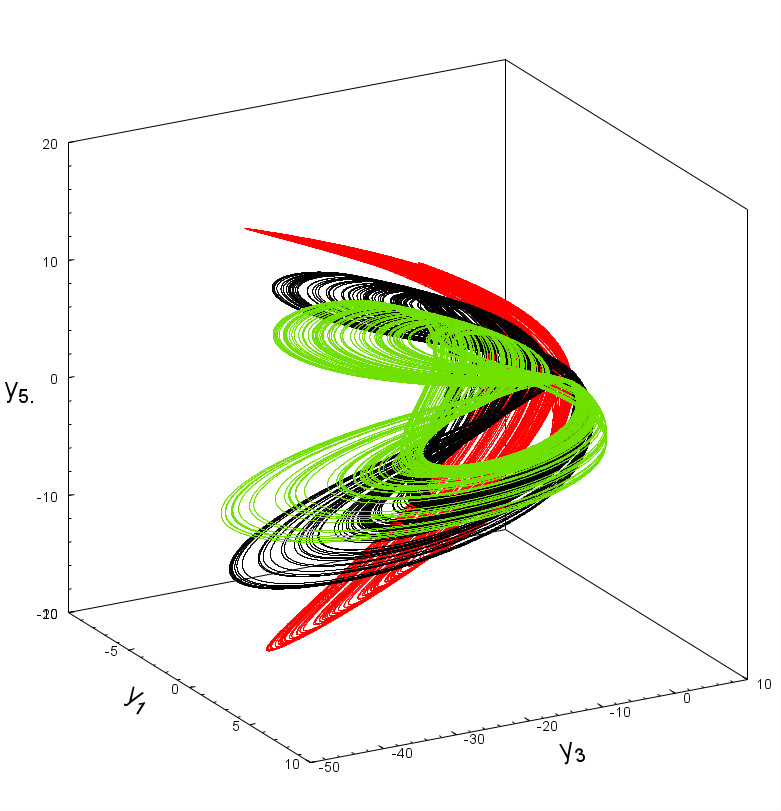}\\
a) $\delta=0$}
\end{minipage}
\hfill
\begin{minipage}[htbp]{0.5\linewidth}
\center{\includegraphics[width=1\linewidth]{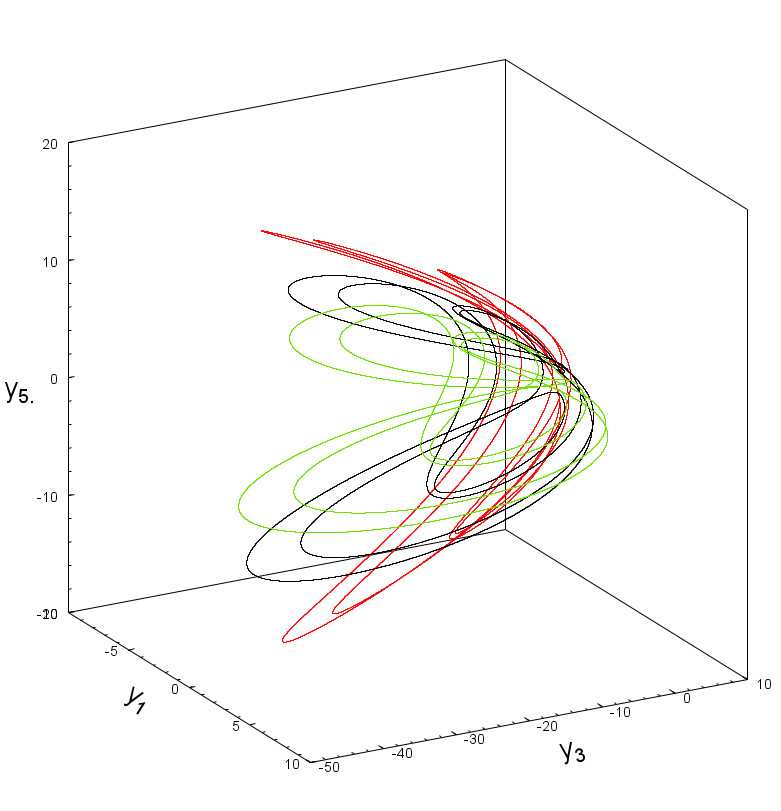}\\
b) $\delta=7\cdot10^{-4}$}
\end{minipage}
\caption{Projections of the phase portraits}
\label{ris:image10}
\end{figure}

In
\cite{shvets2021pendulum,donetskyi2023},
the numerical simulations
were performed
under the assumption
that no time delays
were present
in the
``spherical pendulum--electric motor''
system,
that is,
for
$\delta=\rho=0$.
It was established
that,
under conditions (\ref{par}),
the attractor
is a chaotic
maximal attractor.
In Fig. 2a, projections of the phase portraits of three representatives of the family forming the chaotic maximal attractor are shown in different colors.
All representatives
of the maximal attractor
shown
in Fig.~\ref{ris:image10}a
have the same
positive value
of the largest
Lyapunov characteristic exponent.
This confirms
the chaotic nature
of the attractor.
Moreover,
all representatives
possess the same signature
of the Lyapunov
characteristic exponent (LCE)
spectrum,
namely
$<+,0,0,-,->$.

The presence of time delays
in the
``spherical pendulum--electric motor''
system,
which is equivalent
to nonzero values
of the parameters
$\delta$ and $\rho$
in system (\ref{d}),
may lead
to a substantial change
in the type
of the limiting set.
In particular,
as the time delay
$\delta$
increases,
the chaotic maximal attractor
disappears,
and a new limiting set,
namely
a periodic maximal attractor,
emerges
in system (\ref{d}).
The projections
of the phase portraits
of the corresponding
representatives
of this new type
of maximal attractor
are shown
in Fig.~\ref{ris:image10}b.
The periodic maximal attractor
forms a family
consisting
of infinitely many
closed trajectories
(limit cycles),
all of which
exist simultaneously.
Every neighborhood
of any cycle
contains other cycles
of the family,
that is,
they are not isolated.
However,
these cycles
have neither
points of tangency
nor intersections.
Each closed trajectory
is itself
a limiting set.
This is due
to the fact
that almost every trajectory
originating
from a sufficiently large region
of the phase space
approaches
one of the cycles
of the family.
Nevertheless,
none of these cycles
is an attractor
in the classical sense
of the term.
Therefore,
none of them
is a classical attractor.
It was established
that every cycle
of the family
has the same period.
In addition,
all cycles
have the same signature
of the Lyapunov
characteristic exponent (LCE)
spectrum,
with the largest exponent
being equal to zero
for every cycle.
The analysis
of the Poincaré sections
showed
that all cycles
of the periodic
maximal attractor
possess qualitatively similar
Poincaré sections,
each consisting
of the same finite number
of isolated points.

\begin{figure}[htbp]
\begin{minipage}[htbp]{0.5\linewidth}
\center{\includegraphics[width=1\linewidth]{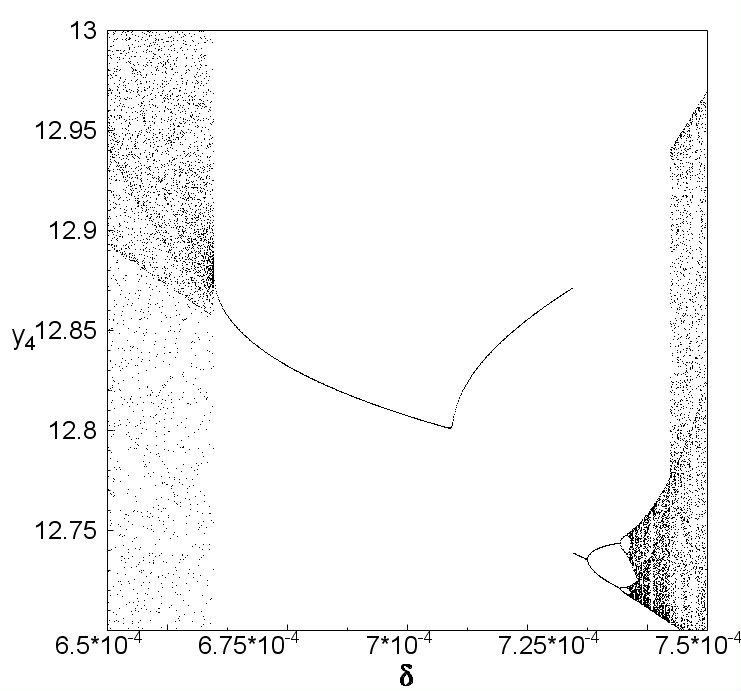} \\ a)}
\end{minipage}
\hfill
\begin{minipage}[htbp]{0.5\linewidth}
\center{\includegraphics[width=1\linewidth]{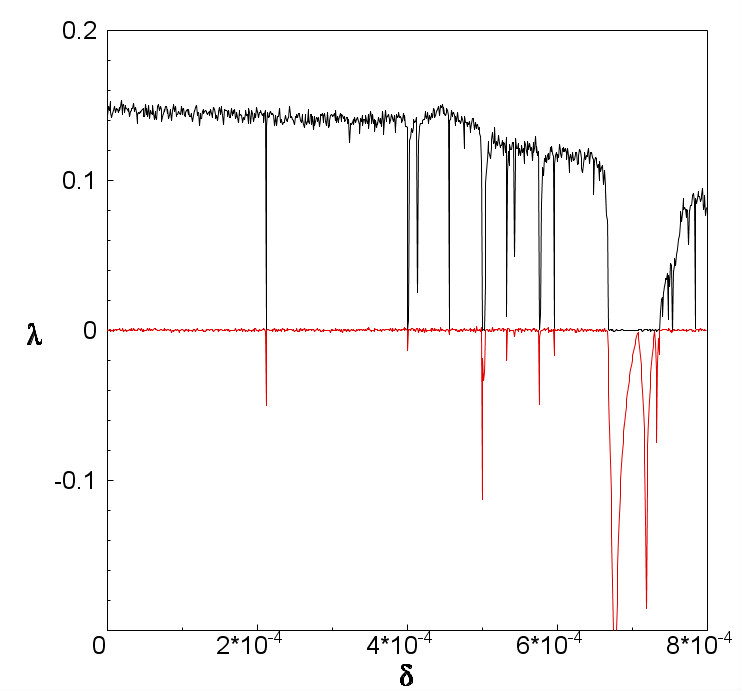} \\ b)}
\end{minipage}
\caption{Phase-parametric characteristic and graphs of two Lyapunov
characteristic exponents}
\label{ris:image8}
\end{figure}

To obtain
a more detailed understanding
of the bifurcations
occurring
in system (\ref{d}),
we also investigate
two other important
characteristics
of this dynamical system,
namely,
its bifurcation diagram
and the dependence
of the Lyapunov
characteristic exponents
on the parameter
$\delta$.

Figure~\ref{ris:image8}a
shows
the bifurcation diagram
of system (\ref{d}).
The densely shaded
black regions
of the diagram
correspond
to chaotic regimes,
whereas
the individual branches
represent
periodic regimes. It should be noted
that the bifurcation diagram
of system (\ref{d})
is constructed
for one representative
of the continuous family
forming the maximal attractor.
In particular,
Fig.~\ref{ris:image8}a
was obtained
using the initial conditions
$y_1=y_2=y_3=y_4=y_5=0.1$.
All bifurcation diagrams
constructed
for different representatives
of the maximal attractor
are qualitatively similar.
They may differ
only in the numerical values
along the vertical axis,
where one
of the phase variables
is plotted.
The pattern
of alternating
densely shaded regions
and individual branches
remains the same,
regardless
of which representative
of the maximal attractor
is used
to construct
the diagram.

This qualitative similarity
of the bifurcation diagrams
makes it possible
to identify
the points
of possible bifurcations,
that is,
transitions
from chaotic
maximal attractors
to periodic
maximal attractors
and vice versa
as the time delays vary.
As can be seen
from Fig.~\ref{ris:image8}a,
an increase
in the parameter
$\delta$
causes
the chaotic regime
to be replaced
by a regular one.
The broad
densely shaded region
of the bifurcation diagram
is replaced
by a single branch,
which corresponds
to the transition
from a chaotic regime
to a periodic one.
With a further increase
in the time delay,
densely shaded regions
reappear
in the bifurcation diagram,
indicating
the emergence
of chaotic
maximal attractors.
For maximal attractors,
as for
``classical''
attractors,
the regions
containing
individual branches
of the bifurcation diagram
are referred to
as periodic windows
within chaos.

The graphs
of the Lyapunov
characteristic exponents
as functions
of the parameter
$\delta$
provide valuable information
for analyzing
the bifurcations
of system (\ref{d}).
Figure~\ref{ris:image8}b
shows
the dependence
of the largest
($\lambda_1$,
black curve)
and the third
($\lambda_3$,
red curve)
Lyapunov
characteristic exponents
on the time delay
$\delta$.
Chaotic
maximal attractors
correspond
to the intervals
along the horizontal axis
for which
the largest
Lyapunov
characteristic exponent
satisfies
$\lambda_1>0$,
whereas
periodic
maximal attractors
correspond
to the intervals
for which
$\lambda_1=0$.
As before,
this graph
was constructed
for a single representative
of the family
forming
the maximal attractor.
The intervals
of chaotic
and periodic dynamics
are identical
for the corresponding graphs
constructed
for any other representative
of the maximal attractor.

The graph
of the third
Lyapunov
characteristic exponent
$\lambda_3$
also provides
important information
about the type
of attractor.
The portions
of this graph
lying
in the negative region
($\lambda<0$)
correspond
to periodic regimes,
whereas
the points
at which the graph
approaches
the line
$\lambda=0$
indicate
bifurcation points.
\begin{figure}[htbp]
\begin{minipage}[htbp]{0.45\linewidth}
\center{\includegraphics[width=1\linewidth]{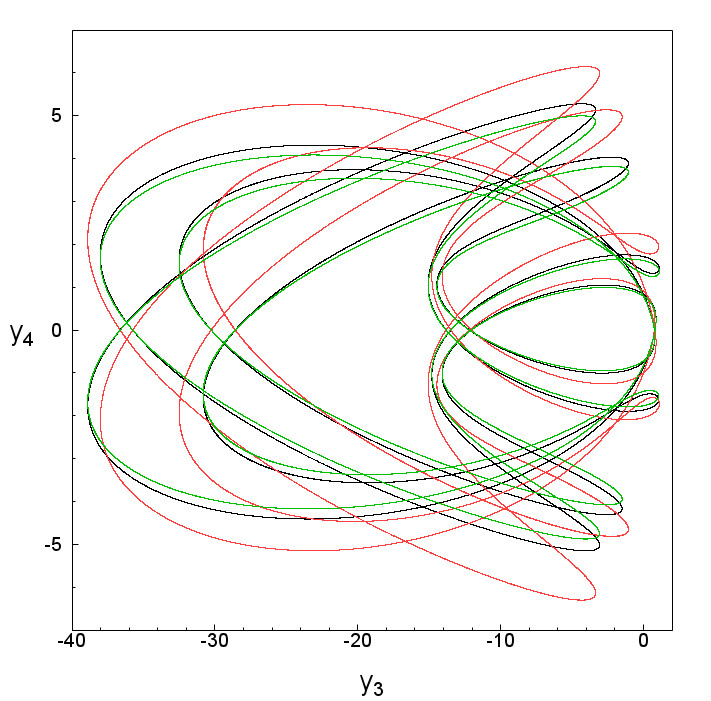}\\
a) $\delta=7.30\cdot10^{-4}$}
\end{minipage}
\hfill
\begin{minipage}[htbp]{0.45\linewidth}
\center{\includegraphics[width=1\linewidth]{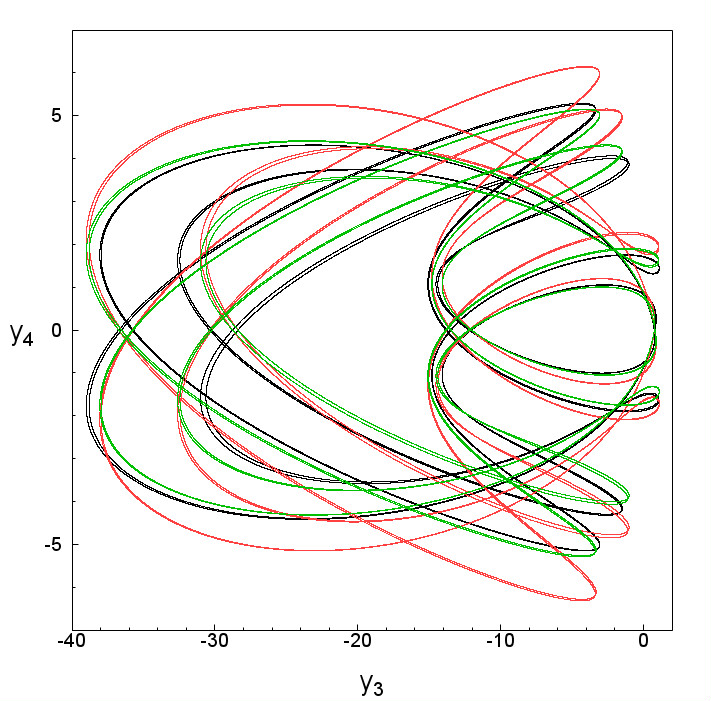}\\
b) $\delta=7.32\cdot10^{-4}$}
\end{minipage}

\vfill

\begin{minipage}[htbp]{0.45\linewidth}
\center{\includegraphics[width=1\linewidth]{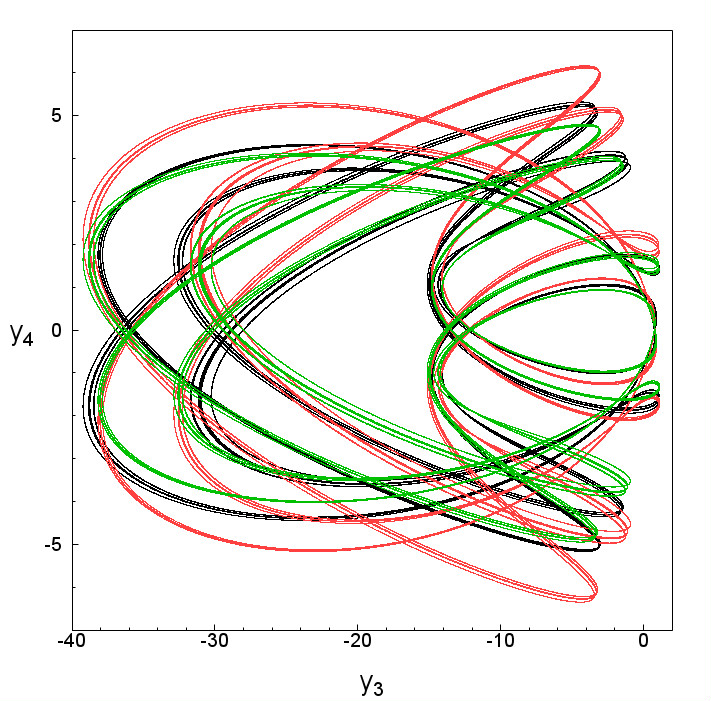}\\
c) $\delta=7.410\cdot10^{-4}$}
\end{minipage}
\hfill
\begin{minipage}[htbp]{0.45\linewidth}
\center{\includegraphics[width=1\linewidth]{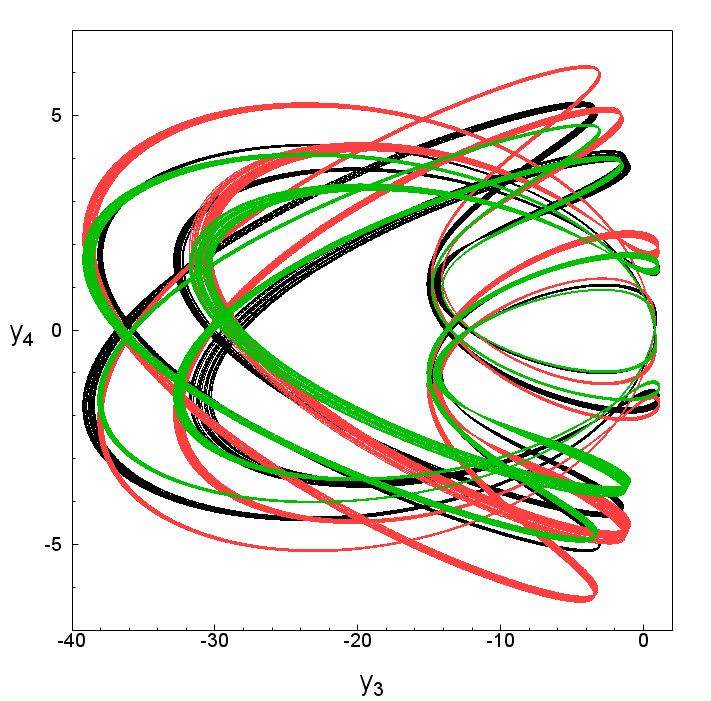}\\
d) $\delta=7.415\cdot10^{-4}$}
\end{minipage}

\caption{Projections of the phase portraits
for the first period-doubling bifurcation
and two chaotic attractors.}
\label{ris:portraits}
\end{figure}

Consider the interval
of time-delay values
$\delta\in(0,\;7.415\cdot10^{-4})$
and examine
the sequence
of bifurcations
observed
within this interval.
For $\delta=0$,
system (\ref{d})
possesses
a chaotic
maximal attractor
of the type
shown
in Fig.~\ref{ris:image10}a.
With a slight increase
in the time delay
to $\delta=7.30\cdot10^{-4}$,
the chaotic
maximal attractor
disappears,
and a regular,
namely
periodic,
maximal attractor
is born
in the system.
The
``chaos--cycle''
transition
occurs
through
a single
hard bifurcation.
For classical attractors,
such a transition
can be interpreted
as the reverse
Pomeau--Manneville scenario
\cite{manneville1980a}.

However,
as has been established
for maximal attractors
(see, for example,
\cite{ShvetsDonetskyi2022,donetskyi2023,Shvets2023_Hydrodynamic}),
the transitions
from regular regimes
to chaotic ones
and vice versa
occur
according
to analogous scenarios.
Figure~\ref{ris:portraits}a
shows,
in different colors,
two-dimensional projections
of three representatives
of the resulting
periodic maximal attractor
for
$\delta=7.30\cdot10^{-4}$.
As noted earlier,
all representatives
of the family
forming the maximal attractor
have the same period.
With a further increase
in the time delay
$\delta$,
each cycle
of the family
undergoes
an infinite cascade
of period-doubling bifurcations.
Moreover,
the period doubling
of all cycles
in the family
takes place
at the same
bifurcation point.

\begin{figure}[htbp]
\begin{minipage}[htbp]{0.45\linewidth}
\center{\includegraphics[width=1\linewidth]{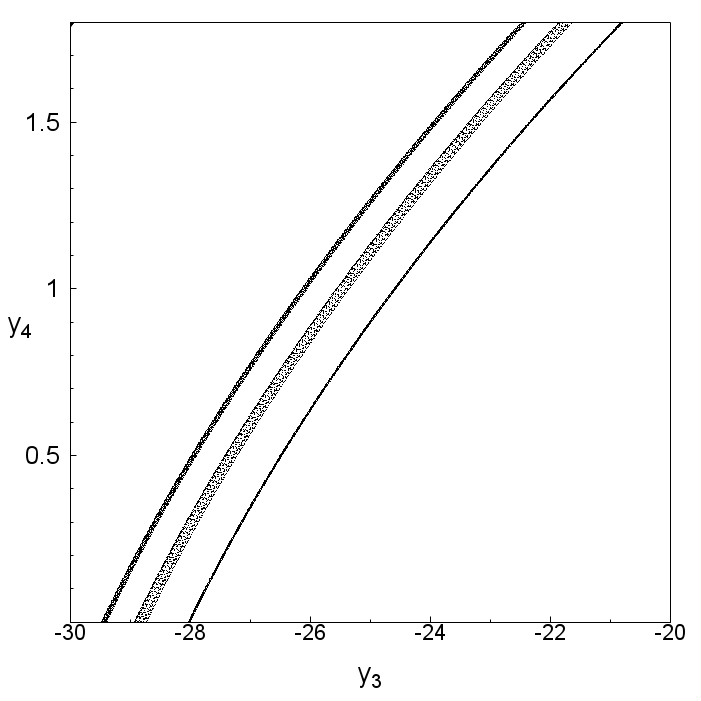}\\
a) $\delta=7.410\cdot10^{-4}$}
\end{minipage}
\hfill
\begin{minipage}[htbp]{0.45\linewidth}
\center{\includegraphics[width=1\linewidth]{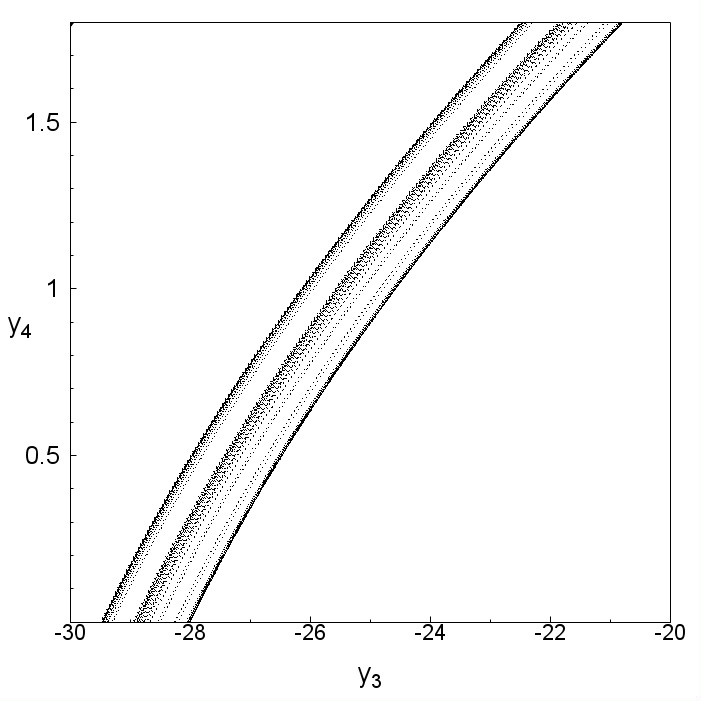}\\
b) $\delta=7.411\cdot10^{-4}$}
\end{minipage}

\vfill

\begin{minipage}[htbp]{0.45\linewidth}
\center{\includegraphics[width=1\linewidth]{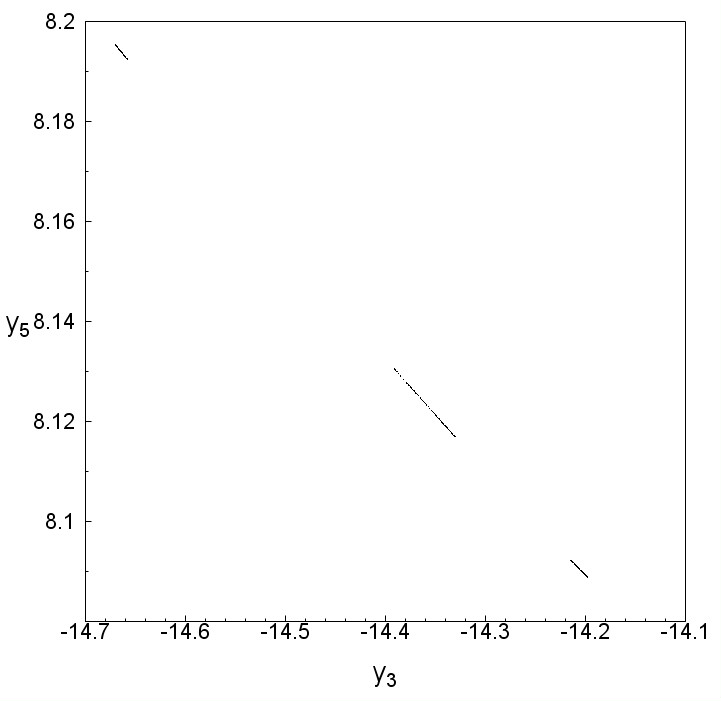}\\
c) $\delta=7.410\cdot10^{-4}$}
\end{minipage}
\hfill
\begin{minipage}[htbp]{0.45\linewidth}
\center{\includegraphics[width=1\linewidth]{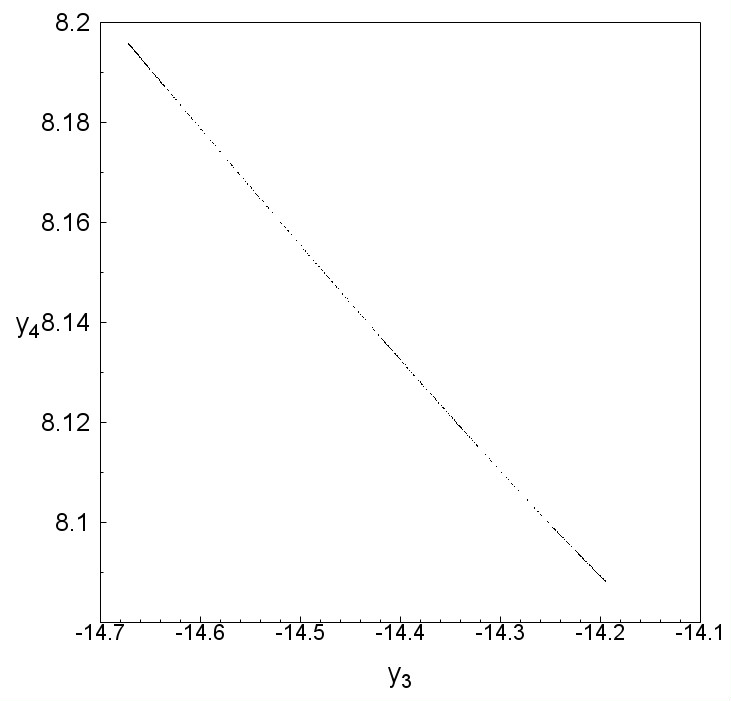}\\
d) $\delta=7.411\cdot10^{-4}$}
\end{minipage}

\caption{Fragments of the natural invariant measure
projected onto one representative
of the maximal attractor
(a, b);
fragments of the Poincaré sections
(c, d).}
\label{ris:experimentalcorrelationsignals11}
\end{figure}

Figure~\ref{ris:image8}a
shows the bifurcation diagram
of one representative
of the maximal attractor.
The period-doubling bifurcations
are represented
by the branching
of individual branches
in this diagram.
For
$\delta=7.30\cdot10^{-4}$,
the system
possesses
a periodic
maximal attractor,
whose three representatives
are shown
in the two-dimensional projection
in Fig.~\ref{ris:portraits}a.
At
$\delta\approx7.21\cdot10^{-4}$,
a period-doubling bifurcation
takes place,
and the period
of every representative
of the periodic
maximal attractor
doubles.
In the graph
of the Lyapunov
characteristic exponents,
this bifurcation
corresponds
to the third exponent
approaching zero.

With a further increase
in the time delay,
the system
undergoes
an infinite sequence
of such
period-doubling bifurcations.
This sequence
is known
as the
Feigenbaum scenario
\cite{feigenbaum1978,feigenbaum1979}.
As a result
of this sequence
of bifurcations,
a chaotic
maximal attractor
emerges
in the system
at
$\delta\approx7.41\cdot10^{-4}$.
Three representatives
of this attractor
are shown
in the three-dimensional projection
in Fig.~\ref{ris:portraits}c.
Figure~\ref{ris:experimentalcorrelationsignals11}c
shows
a fragment
of the Poincaré section
for this attractor.
This Poincaré section
consists
of an infinite number
of points
and possesses
a quasi-ribbon structure.
That is,
the trajectory
evolving
on this attractor
intersects
the Poincaré section
an infinite number
of times,
while
the intersection points
form clusters
whose shape
is close
to a line segment.

With a further increase
in the time delay,
at
$\delta\approx7.411\cdot10^{-4}$,
another important bifurcation
takes place,
namely,
the transition
from a chaotic
maximal attractor
of one type
to a chaotic
maximal attractor
of another type.
This transition
follows
the generalized intermittency
scenario,
which is described
in detail
in the review papers
\cite{shvets2021overview,shvets2025}.
It should be emphasized
that the generalized intermittency
scenario
is realized
both for classical
and for maximal attractors.
For example,
its realization
for classical attractors
of the ideal system
was demonstrated
in \cite{horchakov2026, HorchakovShvets2024SRIT}.

One indication
of this scenario
is a significant expansion
of the densely shaded region
in the bifurcation diagram
after the bifurcation point
is crossed
by the time-delay parameter.
This region
corresponds
to the localization region
of the new type
of chaotic
maximal attractor
in the phase space.
Another characteristic feature
is that
this expansion
of the localization region
is accompanied
by a rapid increase,
almost by a factor of two,
in the largest
Lyapunov
characteristic exponent,
as can be seen
in the graph
shown
in Fig.~\ref{ris:image8}b. 

Figures~\ref{ris:experimentalcorrelationsignals11}a
and~\ref{ris:experimentalcorrelationsignals11}b
show fragments
of the distribution
of the natural invariant measure
over the two-dimensional projections
of the representatives
of the maximal attractors
before and after
the generalized intermittency
bifurcation.
The chaotic
maximal attractor
existing
in system (\ref{d})
for
$\delta<7.41\cdot10^{-4}$
disappears.
Instead,
for
$\delta>7.41\cdot10^{-4}$,
another chaotic
maximal attractor
emerges
(Fig.~\ref{ris:experimentalcorrelationsignals11}b).

The distribution
of the natural invariant measure
over the phase portrait
consists
of two parts:
a densely shaded region,
qualitatively similar
to the attractor
that disappeared
at the bifurcation
(Fig.~\ref{ris:experimentalcorrelationsignals11}a),
and a light-gray region
consisting
of isolated points.
For most of the time,
the representative point
moves
along the part
of the attractor
that resembles
the attractor
destroyed
by the bifurcation.
This is
the coarse-grain (rigid) laminar phase
of generalized intermittency.
However,
at unpredictable moments,
the trajectory
escapes
to more distant regions
of the phase space.
This is
the turbulent phase
of generalized intermittency.
These excursions
correspond
to the isolated points
visible
in the figure.
Such transitions
from the coarse-grain (rigid) laminar phase
to the turbulent phase
and back
occur
an infinite number
of times.
Figures~\ref{ris:experimentalcorrelationsignals11}c
and~\ref{ris:experimentalcorrelationsignals11}d
show fragments
of the Poincaré sections
for one representative
of the maximal attractor.
These sections
also provide
clear evidence
of the realization
of this scenario.
As can be seen,
the Poincaré sections
of the representatives
of this maximal attractor
have a quasi-ribbon structure.
Figures~\ref{ris:experimentalcorrelationsignals11}c
and~\ref{ris:experimentalcorrelationsignals11}d
demonstrate
that the attractor
after the bifurcation
contains
all fragments
of the vanished attractor
shown
in Fig.~\ref{ris:experimentalcorrelationsignals11}c.
These fragments
constitute
the coarse-grain (rigid) laminar phase
of the attractor.

Finally,
we focus
on one point
that requires
additional explanation.
The numerical simulations
have shown
that qualitative changes
in the dynamical behavior
of the considered
non-ideal
``spherical pendulum--electric motor''
system
occur
for very small variations
of the time-delay parameters.
This is
quite natural,
since these delays
are measured
in the
``slow''
time scale.
This time scale
arises
when deriving
the equations of motion
by means
of the averaging method
\cite{Mitropolsky1971}.
The corresponding
physical time,
obtained
by dividing
the slow time
by the small parameter,
is several orders
of magnitude larger.
Incidentally,
the smallness
of the time delays
fully justifies
the reduction
of the delay system
(\ref{c})
to the system
without time delays
(\ref{d}).
The details
of the application
of the averaging method
are presented
in
\cite{krasnopolskaya1992,shvets2007}.

\printbibliography

@article{sommerfeld1902,
  author  = {A. Sommerfeld},
  title   = {Beitr{\"a}ge zum dynamischen Ausbau der Festigkeitslehre},
  journal = {Physikalische Zeitschrift},
  volume  = {3},
  pages   = {266--271},
  year    = {1902}
}

@article{sommerfeld1904,
  author  = {A. Sommerfeld},
  title   = {Beiträge zum dynamischen Ausgleich von Maschinen: II. Die Wirkung eines für unendliche Schwingungen berechneten, aber unvollkommenen Motors auf die Fundamentschwingungen},
  journal = {Zeitschrift des Vereines deutscher Ingenieure},
  volume  = {48},
  number  = {18},
  pages   = {631--636},
  year    = {1904}
}

@book{rocard1943,
  author    = {Y. Rocard},
  title     = {Théorie des oscillations},
  publisher = {Librairie Scientifique A. Hermann et Cie},
  address   = {Paris},
  year      = {1943}
}

@book{kononenko1969,
  author    = {V. O. Kononenko},
  title     = {Vibrating System with a Limited Power-Supply},
  publisher = {Iliffe},
  address   = {London},
  year      = {1969}
}

@article{Balthazar2018Over,
  author  = {J. Balthazar and A. Tusset and R. and R. Brasil and J. Felix and R. Rocha and F. Janzen and A. Nabarrete and C. Oliveira },
  title   = {An overview on the appearance of the Sommerfeld effect and saturation phenomenon in non-ideal vibrating systems (NIS) in macro and MEMS scales},
  journal = {Nonlinear Dyn.},
    volume  = {93},
  pages   = {319--40},
  year    = {2018},
  doi     = {10.1007/s11071-018-4126-0}
}

@article{pintoGenerating2023,
  author  = {J. Pinto and H. Oliveira and A.Cardoso and C.Silva},
  title   = {Generating Wildfire Heat Maps with Twitter and BERT},
  journal = { In: Quaresma, P., Camacho, D., Yin, H., Gonçalves, T., Julian, V., Tallón-Ballesteros, A.J. (eds) Intelligent Data Engineering and Automated Learning – IDEAL 2023. IDEAL 2023. Lecture Notes in Computer Science},
  volume  = {14404},
  pages   = {82--94},
  year    = {2023},
  doi     = {10.1007/978-3-031-48232-8_9}
}

@article{donetskyi2023,
  author  = {S. V. Donetskyi and A. Yu. Svets},
  title   = {Generalization of the concept of attractor for pendulum systems with limited excitation},
  journal = {Journal of Mathematical Sciences},
  volume  = {273},
  number  = {2},
  pages   = {220--229},
  year    = {2023},
  doi     = {10.1007/s10958-023-06550-7}
}

@article{miles1962,
  author  = {J. W. Miles},
  title   = {Stability of forced oscillations of a spherical pendulum},
  journal = {Quart. Appl. Math},
  volume  = {20},
  number  = {1},
  pages   = {21--32},
  year    = {1962},
  doi = {/10.1121/1.1909816}
}

@article{miles1984,
  author  = {J. W. Miles},
  title   = {Resonant motion of a spherical pendulum},
  journal = {Phys. D},
  volume  = {11},
  number  = {3},
  pages   = {309--323},
  year    = {1984},
  doi =     {  10.1016/0167-2789(84)90013-7}
}

@article{miles1984Fara,
  author  = {J. W. Miles},
  title   = {Nonlinear Faraday resonance},
  journal = {J. Fluid Mech.},
  volume  = {146},
  number  = {2},
  pages   = {285--302},
  year    = {1984},
  doi  = {10.1017/S0022112084001865 }
}

@article{krasnopolskaya1992,
  author  = {T. S. Krasnopolskaya and A. Yu. Shvets},
  title   = {Chaotic oscillations of a spherical pendulum as an example of interaction with an energy source},
  journal = {International applied mechanics},
  volume  = {28},
  number  = {10},
  pages   = {52--61},
  year    = {1992},
  doi = {10.1007/BF00846923}
}

@article{shvets2007,
  author  = {A. Yu. Shvets},
  title   = {Deterministic chaos of a spherical pendulum under limited excitation},
  journal = {Ukr. Math. J.},
  volume  = {59},
  number  = {},
  pages   = {602--614},
  year    = {2007},
  doi   =   {10.1007/s11253-007-0039-7}
}

@article{Krasnopolskaya1994Resonance,
  author  = {Krasnopolskaya, T. S. and Podchasov, N. P.},
  title   = {Resonance and Chaos in Nonaxisymmetric Dynamic Processes in Hydroelastic Systems},
  journal = {International Applied Mechanics},
  year    = {1994},
  volume  = {29},
  number  = {12},
  pages   = {1024--1029}
}

@incollection{Shvets2021NewTypes,
  author    = {Shvets, A. Yu. and Donetskyi, S. V.},
  title     = {New Types of Limit Sets in the Dynamic System ``Spherical Pendulum--Electric Motor''},
  booktitle = {Nonlinear Mechanics of Complex Structures},
  editor    = {Altenbach, H. and Amabili, M. and Mikhlin, Y. V.},
  series    = {Advanced Structured Materials},
  volume    = {157},
  pages     = {443--455},
  publisher = {Springer},
  address   = {Cham},
  year      = {2021},
  doi       = {10.1007/978-3-030-75890-5_25},
  
}

@article{Milnor1985Correction,
  author  = {Milnor, John},
  title   = {On the Concept of Attractor: Correction and Remarks},
  journal = {Communications in Mathematical Physics},
  year    = {1985},
  volume  = {102},
  number  = {3},
  pages   = {517--519},
  doi     = {10.1007/BF01209298},
  
}

@book{Anishchenko2014Deterministic,
  author    = {Anishchenko, V. S. and Vadivasova, T. E. and Strelkova, G. I.},
  title     = {Deterministic Nonlinear Systems: A Short Course},
  series    = {Springer Series in Synergetics},
  publisher = {Springer},
  address   = {Cham},
  year      = {2014},
  doi       = {10.1007/978-3-319-06871-8},  
}

@incollection{Cveticanin2018,
  author    = {L. Cveticanin and
               M. Zukovic and
               José Manoel Balthazar},
  title     = {Non-ideal Energy Harvester with Piezoelectric Coupling},
  booktitle = {Dynamics of Mechanical Systems with Non-Ideal Excitation},
  series    = {Mathematical Engineering},
  publisher = {Springer},
  address   = {Cham},
  year      = {2018},
  pages     = {121--144},
  doi       = {10.1007/978-3-319-54169-3_6}
}

@incollection{Warminski2022Nonlinear,
  author    = {Warminski, J.},
  title     = {Nonlinear Dynamics of Self and Parametrically Excited Systems with Non-ideal Energy Source},
  booktitle = {Nonlinear Vibrations Excited by Limited Power Sources},
  series    = {Mechanisms and Machine Science},
  volume    = {116},
  pages     = {53--72},
  publisher = {Springer},
  address   = {Cham},
  year      = {2022},
  doi       = {10.1007/978-3-030-96603-4_5},
  
}

@article{Mikhlin2020Resonance,
  author  = {Mikhlin, Yu. and Onizhuk, A. and Awrejcewicz, J.},
  title   = {Resonance Behavior of the System with a Limited Power Supply Having the Mises Girder as Absorber},
  journal = {Nonlinear Dynamics},
  year    = {2020},
  volume  = {99},
  number  = {1},
  pages   = {519--536},
  doi     = {10.1007/s11071-019-05125-z},  
}

@article{Petrocino2023,
  author  = {Eduardo A. Petrocino and
             José Manoel Balthazar and
             Ângelo M. Tusset and
             others},
  title   = {Dynamic Analysis of an Electromagnetic Vibration Absorber in a Non-Ideal System},
  journal = {Meccanica},
  year    = {2023},
  volume  = {58},
  number  = {2},
  pages   = {287--302},
  doi     = {10.1007/s11012-022-01621-6}
}

@article{Lebedenko2025Stationary,
  author  = {Lebedenko, Ya. O. and Mikhlin, Yu. V.},
  title   = {Stationary Regimes and Transient in Two Systems with Limited Power Supply},
  journal = {Journal of Applied Nonlinear Dynamics},
  year    = {2025},
  volume  = {14},
  number  = {1},
  pages   = {189--210},
  doi     = {10.5890/JAND.2025.03.013 },
}

@book{Magnitskii2006,
  author    = {Magnitskii, N. A. and Sidorov, S. V.},
  title     = {New Methods for Chaotic Dynamics},
  series    = {World Scientific Series on Nonlinear Science, Series A: Monographs and Treatises},
  volume    = {58},
  publisher = {World Scientific},
  address   = {Hackensack, NJ},
  year      = {2006},
  pages     = {xvii + 363},
  isbn      = {978-981-256-817-5},
  doi       = {10.1142/6117},
}

@book{Skiadas2016,
  editor    = {Skiadas, C. H. and Skiadas, Char.},
  title     = {Handbook of Applications of Chaos Theory},
  publisher = {Chapman and Hall/CRC},
  address   = {Boca Raton, FL},
  year      = {2016},
  doi       = {10.1201/b20232},
}

@article{Shvets2023_Hydrodynamic,
author    = {Shvets, A. Y.},
title     = {Nonisolated Limit Sets for Some Hydrodynamic Systems with Limited Excitation},
journal   = {Journal of Mathematical Sciences},
volume    = {274},
pages     = {912--922},
year      = {2023},
publisher = {Springer},
doi       = {10.1007/s10958-023-06650-4},
}

@book{hairer1987,
  author    = {E. Hairer and S. P. N{\o}rsett and G. Wanner},
  title     = {Solving Ordinary Differential Equations I},
  publisher = {Springer},
  year      = {1987},
  doi = {10.1007/978-3-540-78862-1}
}

@article{henon1976,
  author  = {M. H{\'e}non},
  title   = {A two-dimensional mapping with a strange attractor},
  journal = {Communications in Mathematical Physics},
  volume  = {50},
  pages   = {69--77},
  year    = {1976},
  doi     = {10.1007/BF01608556}
}

@article{benettin1976,
  author  = {G. Benettin and L. Galgani },
  title   = {Kolmogorov entropy and numerical experiments},
  journal = {Physical Review A},
  volume  = {14},
  number  = {6},
  pages   = {2338--2342},
  year    = {1976},
  doi     = {10.1103/PhysRevA.14.2338}
}

@article{benettin1980,
  author  = {G. Benettin and L. Galgani},
  title   = {Lyapunov Characteristic Exponents for Smooth Dynamical Systems and for Hamiltonian Systems, a Method for Computing All of Them, Part 2},
  journal = {Meccanica},
  volume  = {15},
  number  = {1},
  pages   = {21--30},
  year    = {1980},
  doi     = {10.1007/BF02128236}
}

@article{manneville1980a,
  author  = {P. Manneville and Y. Pomeau},
  title   = {Different ways to turbulence in dissipative dynamical systems},
  journal = {Physica D: Nonlinear Phenomena},
  volume  = {1},
  number  = {2},
  pages   = {219--226},
  year    = {1980},
  doi     = {10.1016/0167-2789(80)90013-5}
}

@article{feigenbaum1978,
  author  = {M. J. Feigenbaum},
  title   = {Quantitative universality for a class of nonlinear transformations},
  journal = {Journal of Statistical Physics},
  volume  = {19},
  number  = {1},
  pages   = {25--52},
  year    = {1978},
  doi     = {10.1007/BF01020332}
}

@article{feigenbaum1979,
  author  = {M. J. Feigenbaum},
  title   = {The universal metric properties of nonlinear transformations},
  journal = {Journal of Statistical Physics},
  volume  = {21},
  number  = {6},
  pages   = {669--706},
  year    = {1979},
  doi     = {10.1007/BF01107909}
}

@incollection{shvets2021overview,
  author    = {A. Shvets},
  title     = {Overview of Scenarios of Transition to Chaos in Nonideal Dynamic Systems},
  booktitle = {Springer Proceedings in Complexity},
  publisher = {Springer, Cham},
  pages     = {853--864},
  year      = {2021},
  doi       = {10.1007/978-3-030-70795-8_59}
}

@incollection{shvets2025,
  author    = {Aleksandr Shvets},
  title     = {Generalised Intermittency in Non-ideal and ``Classical'' Dynamical Systems},
  editor    = {A. Timokha},
  booktitle = {Analytical and Approximate Methods for Complex Dynamical Systems},
  series    = {Understanding Complex Systems},
  publisher = {Springer},
  address   = {Cham},
  year      = {2025},
  pages     = {75--87},
  doi       = {10.1007/978-3-031-77378-5_5}
}

@incollection{shvets2021pendulum,
  author    = {A. Shvets and S. Donetskyi},
  title     = {New Types of Limit Sets in the Dynamic System "Spherical Pendulum—Electric Motor"},
  booktitle = {Advanced Structured Materials},
  volume    = {157},
  pages     = {443--455},
  year      = {2021},
  publisher = {Springer, Cham},
  doi       = {10.1007/978-3-030-75890-5_25}
}

@article{horchakov2026,
  author  = {O. Horchakov and A. Shvets},
  title   = {Implementation of a generalized intermittency scenario in the Rossler Dynamical System},
  journal = {System Research and Information Technologies},
  volume  = {1},
  pages   = {103--111},
  year    = {2026},
  doi     = {10.20535/SRIT.2308-8893.2026.1.07 }
}

@book{kuznetsov2006,
  author    = {S. P. Kuznetsov},
  title     = {Dynamic Chaos},
  publisher = {Fizmatlit},
  address   = {Moscow},
  year      = {2006},
  pages     = {292}
}

@article{krasnopolskaya1993,
  author  = {T. S. Krasnopolskaya and A. Yu. Shvets},
  title   = {Parametric resonance in the system liquid in tanks + electric motor},
  journal = {Int. Appl. Mech},
  volume  = {29},
  number  = {9},
  pages   = {722--730},
  year    = {1993}
}

@book{Mitropolsky1971,
  author    = {Yu. A. Mitropolsky},
  title     = {Metod usredneniya v nelineinoi mekhanike},
  publisher = {Naukova Dumka},
  address   = {Kyiv},
  year      = {1971},
  pages     = {440}
}

@article{HorchakovShvets2024SRIT,
  author  = {O.O. Horchakov and A.Yu. Shvets},
  title   = {Generalized Scenarios of Transition to Chaos in Ideal Dynamic Systems},
  journal = {System Research and Information Technologies},
  year    = {2024},
  number  = {3},
  pages   = {64--73},
  doi     = {10.20535/SRIT.2308-8893.2024.3.04}
}

@incollection{ShvetsDonetskyi2022_1,
  author    = {Shvets, A. and Donetskyi, S.},
  title     = {Maximal Attractors in Nonideal Hydrodynamic Systems},
  booktitle = {14th Chaotic Modeling and Simulation International Conference. CHAOS 2021},
  editor    = {Skiadas, C.H. and Dimotikalis, Y.},
  series    = {Springer Proceedings in Complexity},
  publisher = {Springer},
  address   = {Cham},
  year      = {2022},
  pages     = {433--443},
  doi       = {10.1007/978-3-030-96964-6_31}
}

@article{SeitDzhelilShvets2026,
  author        = {Seit-Dzhelil, I. and Shvets, A.},
  title         = {Delay Factors and Genesis of the Limit Sets of a Nonideal ``Tank with Liquid--Electric Motor'' System},
  journal       = {Ukrainian Mathematical Journal},
  year          = {2026},
  volume        = {78},
  articlenumber = {56},
  doi           = {10.1007/s11253-026-02609-1}
}

@article{SeitDzhelilShvets2026Delay,
  author        = {Seit-Dzhelil, I. and Shvets, A.},
  title         = {Influence of Delay on the Regular and Chaotic Dynamics of a ``Tank with Liquid--Electric Motor'' System},
  journal       = {Ukrainian Mathematical Journal},
  year          = {2026},
  volume        = {78},
  articlenumber = {34},
  doi           = {10.1007/s11253-026-02585-6}
}

@inproceedings{KrasnopolskayaShvets1991,
  author    = {T.S. Krasnopolskaya and A.Yu. Shvets},
  title     = {Chaos in Dynamics of Machines with a Limited Power-Supply},
  booktitle = {Proceedings of the 8th World Congress on the Theory of Machines and Mechanisms},
  editor    = {M. Okrolnick and L. Pust},
  publisher = {Czechoslovak Academy of Sciences},
  address   = {Prague},
  year      = {1991},
  volume    = {1},
  pages     = {181--184}
}

@article{ShvetsMakaseyev2012,
  author  = {Shvets, A.Yu. and Makaseyev, A.M.},
  title   = {Chaotic Oscillations of Nonideal Plane Pendulum Systems},
  journal = {Chaotic Modeling and Simulation (CMSIM) Journal},
  year    = {2012},
  number  = {1},
  pages   = {195--204}
}

@article{ShvetsDonetskyi2022,
  author  = {Donetskyi, S.V. and Shvets, A.Yu.},
  title   = {Bifurcations of Maximal Attractors in Nonideal Pendulum Systems},
  journal = {Dopovidi of the National Academy of Sciences of Ukraine},
  year    = {2022},
  number  = {3},
  pages   = {13--19},
  doi     = {10.15407/dopovidi2022.03.013}
}

@incollection{Mitropolsky1980,
  author    = {Mitropolsky, Yu. A. and Shvets, A. Yu.},
  title     = {On the Influence of Delay on the Stability of a Pendulum with a Vibrating Suspension Point},
  booktitle = {Analytical Methods for the Investigation of Nonlinear Oscillations},
  publisher = {Institute of Mathematics, Academy of Sciences of the Ukrainian SSR},
  address   = {Kyiv},
  year      = {1980},
  pages     = {115--120},
}

\end{document}